\documentclass[iop]{emulateapj}
\usepackage{graphicx}
\usepackage{amsmath}
\usepackage{natbib}
\usepackage{longtable}

\shorttitle{Lagrange Points in BH Binaries}
\shortauthors{Schnittman}

\begin{document}

\title{The Lagrange Equilibrium Points $L_4$ and $L_5$ in a Black Hole
Binary System}
\author{Jeremy D.\ Schnittman}
\affil{Department of Physics and Astronomy, Johns Hopkins University, Baltimore, MD 21218}

\begin{abstract}
We calculate the location and stability of the $L_4$ and $L_5$
Lagrange equilibrium points in the circular restricted three-body
problem as the binary system evolves via gravitational radiation
losses. Relative to the purely Newtonian case, we find that the $L_4$
equilibrium point moves towards the secondary mass and becomes slightly
less stable, while the $L_5$ point moves away from the secondary and
gains in stability. We discuss a number of astrophysical applications
of these results, in particular as a mechanism for producing
electromagnetic counterparts to gravitational-wave signals. 
\end{abstract}


\maketitle

%--------------------------------------------------------
\section{INTRODUCTION}
\label{intro}
%--------------------------------------------------------

Recent advances in numerical relativity (NR) have led to an increasing
interest in the astrophysical implications of black hole (BH)
mergers. Of particular interest is the possibility of a distinct,
luminous electromagnetic (EM) counterpart to a gravitational wave (GW)
signal \citep{bloom:09}. If such an EM counterpart could be identified
with a LISA\footnote{\tt http://lisa.nasa.gov}
detection of a supermassive BH binary in the merging process, then the
host galaxy could likely be determined. A large variety of potential
EM signatures have recently been proposed, almost all of which require
some significant amount of gas in the near vicinity of the merging
BHs.

Gas in the form of accretion disks around single massive BHs is known to
produce some of the most luminous objects in the universe. However,
very little is known about the behavior of accretion disks around {\it
  two} BHs, particularly at late times in their inspiral evolution. In
Newtonian disks, it is believed that a circumbinary accretion disk
will have a central gap of much lower density, either
preventing accretion altogether, or at least decreasing it
significantly \citep{pringle:91, artymowicz:94, artymowicz:96}. When
including the evolution of the binary due
to GW losses, the BHs may also decouple from the disk at the point
when the GW inspiral time becomes shorter than the gaseous inflow time at
the inner edge of the disk \citep{milos:05}. This decoupling should
effectively stop accretion onto the central object until the gap can
be filled on an inflow timescale. However, other semi-analytic
calculations predict an enhancement of accretion power as the evolving
binary squeezes the gas around the primary BH, leading to a rapid
increase in luminosity shortly before merger
\citep{armitage:02,chang:09}. 

Setting aside for the moment the question of {\it how} the gas can or
cannot reach the central BH region, a number of recent papers have
shown that {\it if} there is sufficient gas present, then an observable
EM signal is likely. \citet{krolik:10} used analytic arguments to
estimate a peak luminosity comparable to that of the Eddington limit,
independent of the detailed mechanisms for shocking and heating the
gas. Using relativistic magneto-hydrodynamic simulations in 2D,
\citet{oneill:09} showed that the prompt mass loss due to GWs actually
leads to a sudden {\it decrease} in luminosity following the merger, as the
gas in the inner disk temporarily has too much energy and angular momentum to
accrete efficiently. Full NR simulations of the final few orbits of a
merging BH binary have now been carried out including the presence of
EM fields in a vacuum \citep{palenzuela:09, mosta:10, palenzuela:10}
and also gas, treated as test particles in \citet{vanmeter:10} and as
an ideal fluid in \citet{bode:10} and \citet{farris:10}. The
simulations including matter all suggest that the gas can get shocked
and heated to high temperatures, thus leading to bright counterparts
in the event that sufficient gas is in fact present. One estimate of
``sufficient'' is the condition that the gas be optically thick to
electron scattering, which will lead to Eddington-rate luminosities
\citep{krolik:10}. Thus we require $\kappa \rho R \gtrsim 1$, where
$\kappa=0.4$ cm$^2$g$^{-1}$ is the opacity and $\rho$ is the characteristic
density over a length scale $R\sim 10\, GM_{\rm BH}/c^2$. Then the total
mass is a modest $M_{\rm gas}\sim \rho R^3 \approx 5\times 10^{26}\, M_7^2$ g,
where $M_7=M_{\rm BH}/(10^7 M_\odot)$. More gas will generally lead to
longer, but not necessarily more luminous signals.

The motivation for this work is to understand potential dynamical
processes that could lead to a significant amount of gas around the
BHs at the time of merger. We focus on stable orbits in the circular,
restricted three-body problem, i.e., a system with two massive
objects orbiting each other on circular, planar Keplerian orbits,
with the third body being a test particle free to move out of the
plane. Gas or stars that get captured into these stable regions at
large binary separation may remain trapped as the binary orbit shrinks
due to radiation reaction. In this paper, we integrate the restricted
three-body equations of motion in the corotating frame for a range of
BH mass ratios, including the evolution of the binary with
leading-order post-Newtonian GW losses. With the exception of these
non-conservative corrections, all dynamics are considered at the
purely Newtonian level. 

We find that particles initially close to the $L_4$ and $L_5$ Lagrange
points remain on stable orbits throughout the adiabatic evolution of
the BH binary. This result immediately leads to
a number of astrophysical predictions. If diffuse gas is captured from
the surrounding medium, and irradiated by an accretion disk around
either BH, it may produce strong, relatively narrow emission lines
with large velocity
shifts, varying on an orbital time scale. If individual stars are
captured, they may eventually get tidally disrupted by one of the BHs
shortly before merger, providing an enormous source of material for
accretion. Stars may also get ejected from the system at late times, producing
ultra-hyper-velocity stars moving away from the galactic center at a
significant fraction of the speed of light. Clusters of stars could
also get captured or form {\it in situ} at large binary separation,
then get compressed adiabatically as the BH system shrinks, ultimately
collapsing from gravitational instabilities. Compact objects such as
neutron stars or stellar-mass BHs would not get tidally disrupted in most
cases, but could still possibly be detected with LISA as perturbations
to the GW signal.

While this manuscript was in preparation, a similar work by
\citet{seto:10} was published, with some of the same conclusions. One
major distinction is that they included first-order post-Newtonian
terms in their equations of motion, while we do not. Also, they ``have
not found notable qualitative differences'' between $L_4$ and $L_5$,
while we do. Many of our astronomical predictions are qualitatively
similar, but have been derived independently. 

The outline for this paper is as follows: in Section
\ref{location_stability} we present the three-body
equations of motion, including GW losses, and analyze the stability
and location of the equilibrium Lagrange points. These analytic
predictions are tested in Section \ref{numerical_tests} with numerical
simulations of a large number of test particles. In Section
\ref{applications} we explore in more detail some potential
astronomical applications of these results, and in Section
\ref{discussion} we conclude. 

%--------------------------------------------------------
\section{LOCATION AND EVOLUTION OF STABLE LAGRANGE POINTS}
\label{location_stability}
%--------------------------------------------------------

We consider a three-body system wherein the larger bodies have masses
$M_1$ and $M_2$ ($M_1>M_2$; $M\equiv M_1+M_2)$ and the third object is
a test particle 
of negligible mass $m_0$. We define dimensionless masses $\mu_1\equiv M_1/M$
and $\mu_2\equiv M_2/M$. $M_1$ and $M_2$ move on circular,
Keplerian orbits around the center of mass at the origin with
semi-major axis $a(t)$. Unless otherwise stated, we use geometrized units with
$G=M=c=1$. In the corotating frame we define coordinates $(x,y,z)$,
with the primary BH located at $(-\mu_2\, a,
0, 0)$ and the secondary at $(\mu_1\, a,0,0)$. The angular velocity
of the system in the lab frame is $n=a^{-3/2}$. The binary is orbiting
in the $+\phi$ direction for the standard definition of spherical
coordinates. The equations of motion for the test particle may be written as
\begin{subequations}
\begin{eqnarray}
\label{eoma}
\ddot{x} &=& 2n\dot{y}+n^2x +\dot{n}y-
\left[\mu_1\frac{x+a\mu_2}{r_1^3}+\mu_2\frac{x-a\mu_1}{r_2^3}\right] \\
\label{eomb}
\ddot{y} &=& -2n\dot{x}+n^2y -\dot{n}x-
\left[\frac{\mu_1}{r_1^3}+\frac{\mu_2}{r_2^3}\right]y \\
\label{eomc}
\ddot{z} &=& -\left[\frac{\mu_1}{r_1^3}+\frac{\mu_2}{r_2^3}\right]z \\
\label{eomd}
\dot{n} &=& \frac{96}{5}\mu_1\mu_2 n^{11/3} \, ,
\end{eqnarray}
\end{subequations}
where $r_1$ and $r_2$ are the distances to the primary and secondary
bodies, respectively.
These reproduce the classical Newtonian system in the limit of
$\dot{n} \to 0$, i.e.\ when gravitational radiation reaction is turned
off. 

\begin{figure}
\begin{center} 
\scalebox{0.6}{\includegraphics{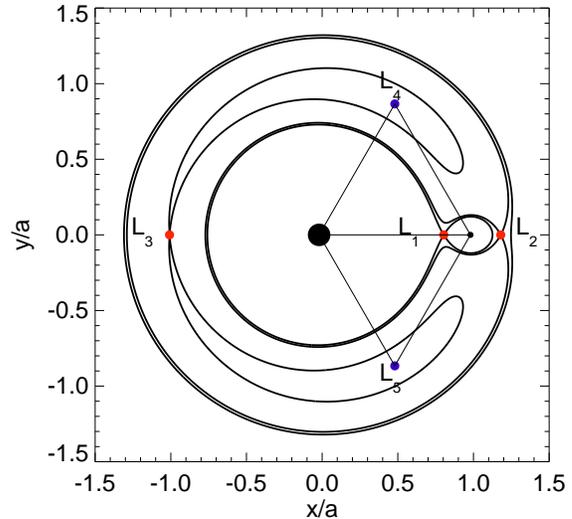}}
\caption{\label{cj_cont} Contours of the Jacobi constant $C_{\rm J0}$ for
  test particles at rest in the rotating frame, for secondary mass
  $\mu_2=0.02$. The stable Lagrange points $L_4$ and $L_5$ are located
  at local minima of $C_{\rm J0}$, and form equilateral triangles with
  the two massive bodies. The co-linear Lagrange points $L_1$, $L_2$,
  and $L_3$ are located at saddle-points of $C_{\rm J0}$ and are
  unstable equilibrium points. The $x-$ and $y-$axes have been
  normalized to the binary separation $a$.} 
\end{center}
\end{figure}

It is well-known from classical mechanics that there exist five
equilibrium points where a test particle can remain at rest in the
corotating frame ($\ddot{x}=\ddot{y}=\ddot{z}=0$)
\citep{murray:99}. These points are known as the Lagrange 
equilibrium points $(L_1 \ldots L_5)$, and are plotted in Figure \ref{cj_cont} for a
binary with mass ratio $\mu_2=0.02$. The co-linear points $L_1$,
$L_2$, and $L_3$ are always unstable, while the triangular points
$L_4$ and $L_5$ are linearly stable only for $\mu_2\lesssim 0.0385$
\citep{murray:99}. In the Newtonian problem, there is a single
integral of motion, the Jacobi constant:
\begin{equation}\label{C_J}
C_{\rm J} \equiv n^2(x^2+y^2) + 2\left(\frac{\mu_1}{r_1}+\frac{\mu_2}{r_2}\right)
-\dot{x}^2-\dot{y}^2-\dot{z}^2.
\end{equation}
Setting the particle's velocity to zero in the corotating frame, the
contours of constant $C_{{\rm J}0}\equiv C_{\rm J}(v=0)$ in the $x-y$ plane define ``zero velocity curves,'' as plotted in
Figure \ref{cj_cont}. $L_4$ and $L_5$ are located at local minima of
$C_{{\rm J}0}$, while $L_1$, $L_2$, and $L_3$ are saddle-points (note that
$C_{\rm J}$ has the form of a negative energy, so that local minima are
actually points of maximum potential energy, yet are still stable to
small perturbations because of the restoring Coriolis forces in the
rotating frame). One well-known
implication of this feature is that gas bound to either $M_1$ or $M_2$
can pass through $L_1$ and subsequently accrete onto the other object,
a process known as mass transfer in standard stellar binary
evolution. Additionally, any particle with $C_{\rm J}>C_{\rm J0}(L_3)$ would be
excluded from crossing into the tadpole-shaped regions around $L_4$
and $L_5$. 

\citet{murray:94} explore the effects of generalized non-gravitational
forces acting on test particles in the restricted three-body problem,
including a detailed study of how the position and stability of the
Lagrange equilibrium points are affected. At first glance, our problem
would seem quite similar, with the $\dot{n}$ terms in equations
(\ref{eoma}),(\ref{eomb}) acting as psuedo-drag forces: as the binary
orbit shrinks and $n$ increases, test particles would feel like they
are accelerating in the $-\phi$ direction in the corotating frame,
thus slowing down in the inertial frame. In this case,
\citet{murray:94} show that $L_4$ and $L_5$ both move in the $+\phi$
direction, thus compensating for the inertial drag and still
maintaining $\ddot{x}=\ddot{y}=0$. However, when considering the
time-varying nature of $n$ and $a$ in equations
(\ref{eoma})-(\ref{eomd}) due to GW losses, we find that this drag force
analogy breaks down. Numerical integration of equations
(\ref{eoma})-(\ref{eomd}) show that in fact, the opposite occurs, with
$L_4$ and $L_5$ moving in the $-\phi$ direction in the rotating frame. 

Remarkably, this result can still be understood via an inertial drag
term, if one applies it not to the test particle, but to the secondary
mass. In order to study the evolution of the Trojan asteroids under
the influence of a radial migration by Jupiter, \citet{fleming:00}
apply an artificial drag force that acts only on Jupiter in the form
$\mathbf{F}=-k\mathbf{v}$ in the inertial frame. Of course, this
causes the secondary to lose energy and angular momentum and thus
orbit faster, increasing $n$ as in the BH binary case. Since there is
no requirement on the relative masses of the secondary $M_2$ and the
test particle $m_0$ to ensure
stability of $L_4$ and $L_5$ \citep{salo:88}, we can in fact adopt the
drag force analysis of \citet{murray:94} to include GW losses by
reversing the roles of the secondary BH and the test particle. In the
rotating frame, the inertial drag force is given by
$F_x=-k(\dot{x}-ny)$ and $F_y=-k(\dot{y}+nx)$. For stationary
particles at the 
equilibrium points, equations (\ref{eoma}),(\ref{eomb}) become
\begin{subequations}
\begin{eqnarray}
\label{eom2a}
\bar{k}ny &=& -n^2x 
+\left[\mu_1\frac{x+a\mu_2}{r_1^3}+\mu_2\frac{x-a\mu_1}{r_2^3}\right] \\
\label{eom2b}
\bar{k}nx &=& n^2y
-\left[\frac{\mu_1}{r_1^3}+\frac{\mu_2}{r_2^3}\right]y \, ,
\end{eqnarray}
\end{subequations}
where $\bar{k}=k/m_0$, and we have replaced the GW acceleration terms
$\dot{n}y$ and $\dot{n}x$ with those appropriate for drag forces
$F_x/m_0$ and $F_y/m_0$. 

In order to reproduce the GW evolution of equation (\ref{eomd}), we must
choose the magnitude of the drag force appropriately. This is done by
equating the energy loss due to the shrinking orbit with that of the
inertial drag force. For circular Keplerian orbits, we have
\begin{equation}\label{edot_drag}
\dot{E}_{\rm drag}=\mathbf{F}\cdot\mathbf{v}=-kv^2=-kn^{2/3}
\end{equation}
and
\begin{equation}\label{edot_gw}
\dot{E}_{\rm orbit}=\frac{d}{dt}\left(-\frac{m_0}{2a}\right)=
-\frac{m_0}{3}n^{-1/3}\dot{n} \, .
\end{equation}
Equating (\ref{edot_drag}) and (\ref{edot_gw}), along with
(\ref{eomd}) allows us to solve for $\bar{k}$:
\begin{equation}\label{kbar_gw}
\bar{k} = \frac{1}{3} \frac{\dot{n}}{n}=\frac{32}{5}\mu_1 \mu_2
n^{8/3}. 
\end{equation}

\begin{figure}
\begin{center} 
\scalebox{0.45}{\includegraphics{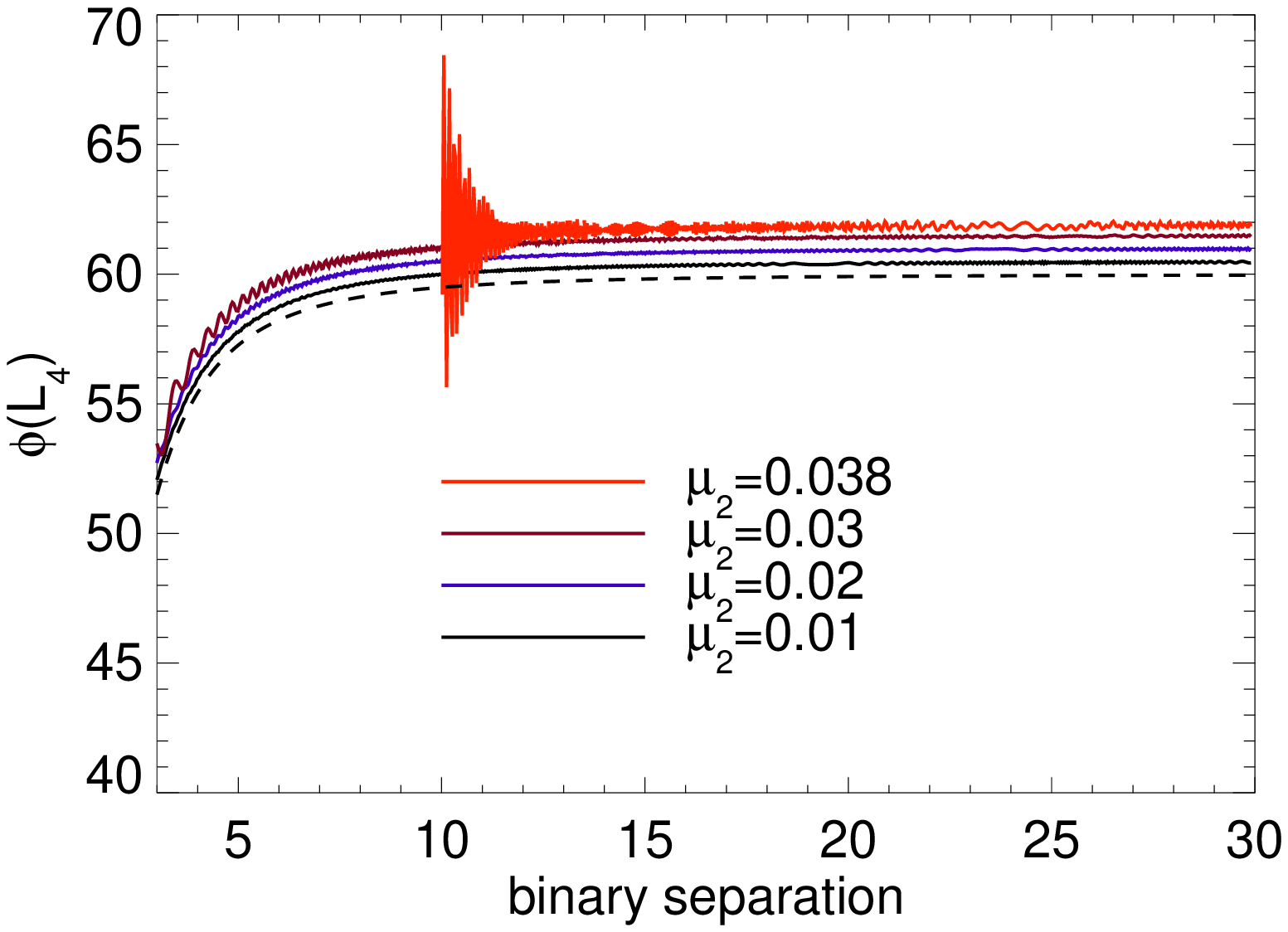}}\\
\scalebox{0.45}{\includegraphics{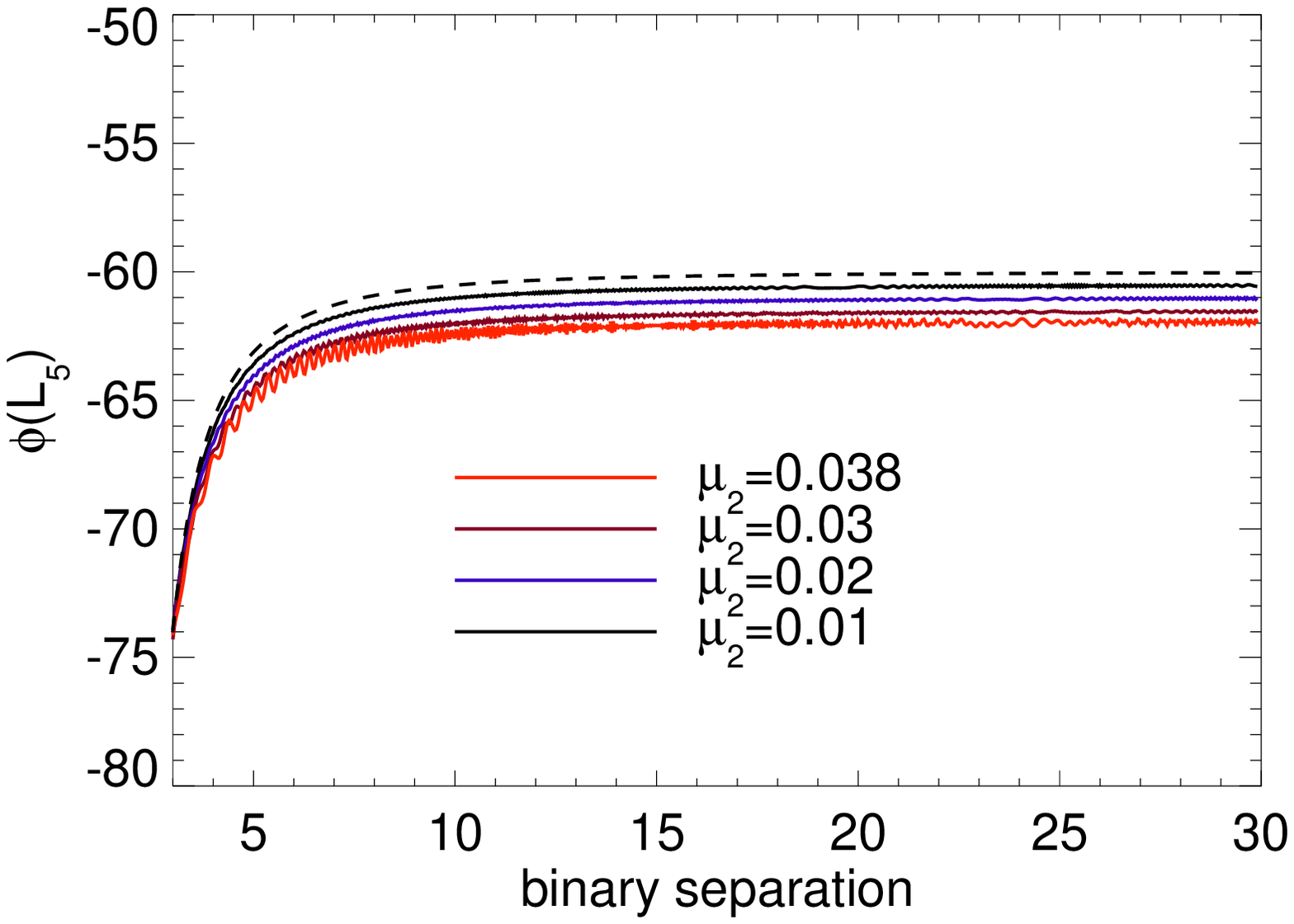}}
\caption{\label{dphi} \textit{Upper:} The location of test
  particles initially at rest near the $L_4$ Lagrange point in the
  rotating frame. As the binary system evolves to smaller separation
  $a$, the $L_4$ point moves towards the secondary BH, located at
  $\phi=0$. For $\mu_2=0.038$, $L_4$ becomes linearly unstable around
  $a=15M$, and the test particle is ejected at $a=10M$. 
  \textit{Lower:} The location of test particles initially
  at rest near the $L_5$ Lagrange point. Towards the end of the BH
  inspiral, the $L_5$ point moves {\it away} from the secondary. In
  both figures, we plot the analytic prediction of equation
  (\ref{phi_L}) as a dashed curve.}
\end{center}
\end{figure}

Following \citet{murray:94}, we multiply equation (\ref{eom2a}) by $y$
and add to (\ref{eom2b}) times $x$, giving 
\begin{equation}\label{kbar_lagrange}
\bar{k}nr^2 = \mu_1\mu_2 a y \left(\frac{1}{r_1^3}-\frac{1}{r_2^3}\right).
\end{equation}
In the limit of small $\mu_2$, $a\approx r \approx r_1$ at the stable Lagrange
points and we can write 
\begin{equation}\label{kbar_lagrange2}
\bar{k} \approx \frac{\mu_1 \mu_2}{na^3}\,
\sin{\phi_L}\left(1-\frac{1}{(2-2\cos\phi_L)^{3/2}}\right),
\end{equation}
where $\phi_L$ is the location of the Lagrange point in the presence
of an inertial drag force. When considering the orbital evolution due
to GW losses, we can use equation (\ref{kbar_gw}) for $\bar{k}$. We
must also reverse the sign of $\phi_L$ because, as mentioned above,
the GW drag force is really applied to $M_2$ and not to the test
particle as in \citet{murray:94}. We now have an expression for the
location of the Lagrange points in the presence of GW losses:
\begin{equation}\label{phi_L}
\sin{\phi_L}\left(\frac{1}{(2-2\cos\phi_L)^{3/2}}-1\right) =
\frac{32}{5}a^{-5/2}.
\end{equation}
This relation is plotted as a dashed curve in Figure \ref{dphi}. 
Interestingly, to leading order, the location of the Lagrange points is
a function only of the binary orbital separation, {\it not} the mass
ratio, even though the inspiral timescale is strongly dependent on
$\mu_2$. Note that, when including non-conservative forces such as
GW evolution, there is a lack of symmetry between $L_4$ and $L_5$,
with $L_4$ moving towards the secondary mass, while $L_5$ moves away
from it. We also find differences in the relative stability of orbits
around each Lagrange point, as described in the next section. 

%--------------------------------------------------------
\section{NUMERICAL TESTS}
\label{numerical_tests}
%--------------------------------------------------------

In Figure \ref{dphi} we plot the position of test particles near $L_4$
and $L_5$ as a
function of the binary separation $a$ for a range of mass ratios. The
initial binary separation is $a_0=40M$. For the most part, the test
particles remain very close to their initial locations throughout
inspiral, evolving according to equation (\ref{phi_L}). While our
Newtonian equations of motion do not formally predict an inner-most
stable circular orbit (ISCO), we expect the BHs to plunge and merge
around $a=5M$, at which point very little evolution in $\phi_L$ has
taken place. The slight offsets in the various curves at large $a$
are due to the fact that $\phi$ is measured with respect to the
origin, not $M_1$.
The inclusion of post-Newtonian (PN) corrections to the
equations of motion would likely change the shape of the curves in
Figure \ref{dphi} slightly, as well as the detailed shape of stability
regions around $L_4$ and $L_5$ \citep{rosswog:96}.  It should be
noted that \citet{seto:10} include only 1PN terms, and do find a
significant reduction in stability for certain mass ratios. 
Yet due to the notoriously uneven
convergence of most PN expansions (e.g.\ \citet{buonanno:09} and
references therein), it is not clear whether their inclusion
would add significant physical insight to the problem.

\begin{figure*}
\begin{center}
\scalebox{0.37}{\includegraphics*[55,425][540,700]{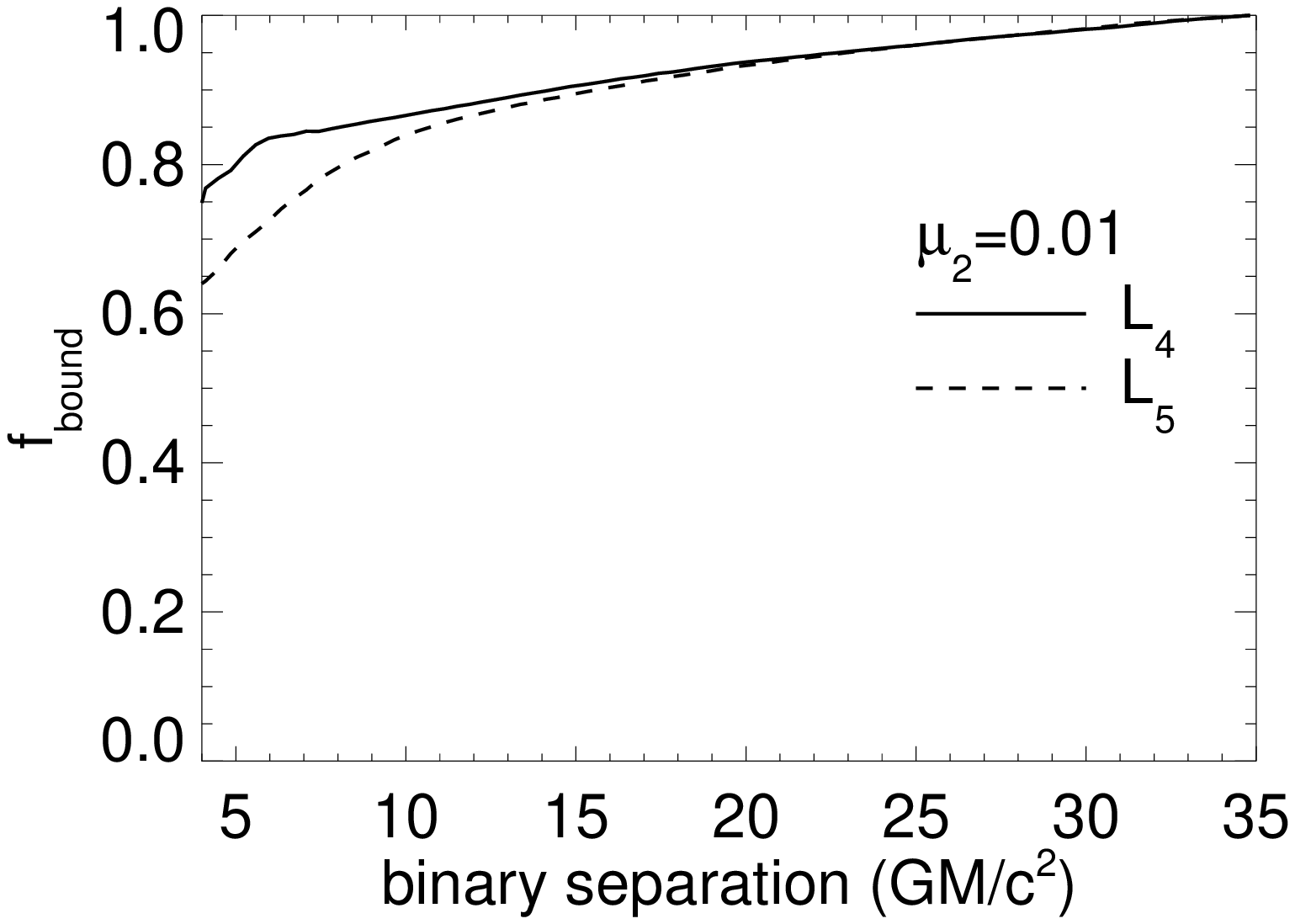}}
\scalebox{0.37}{\includegraphics*[160,425][540,700]{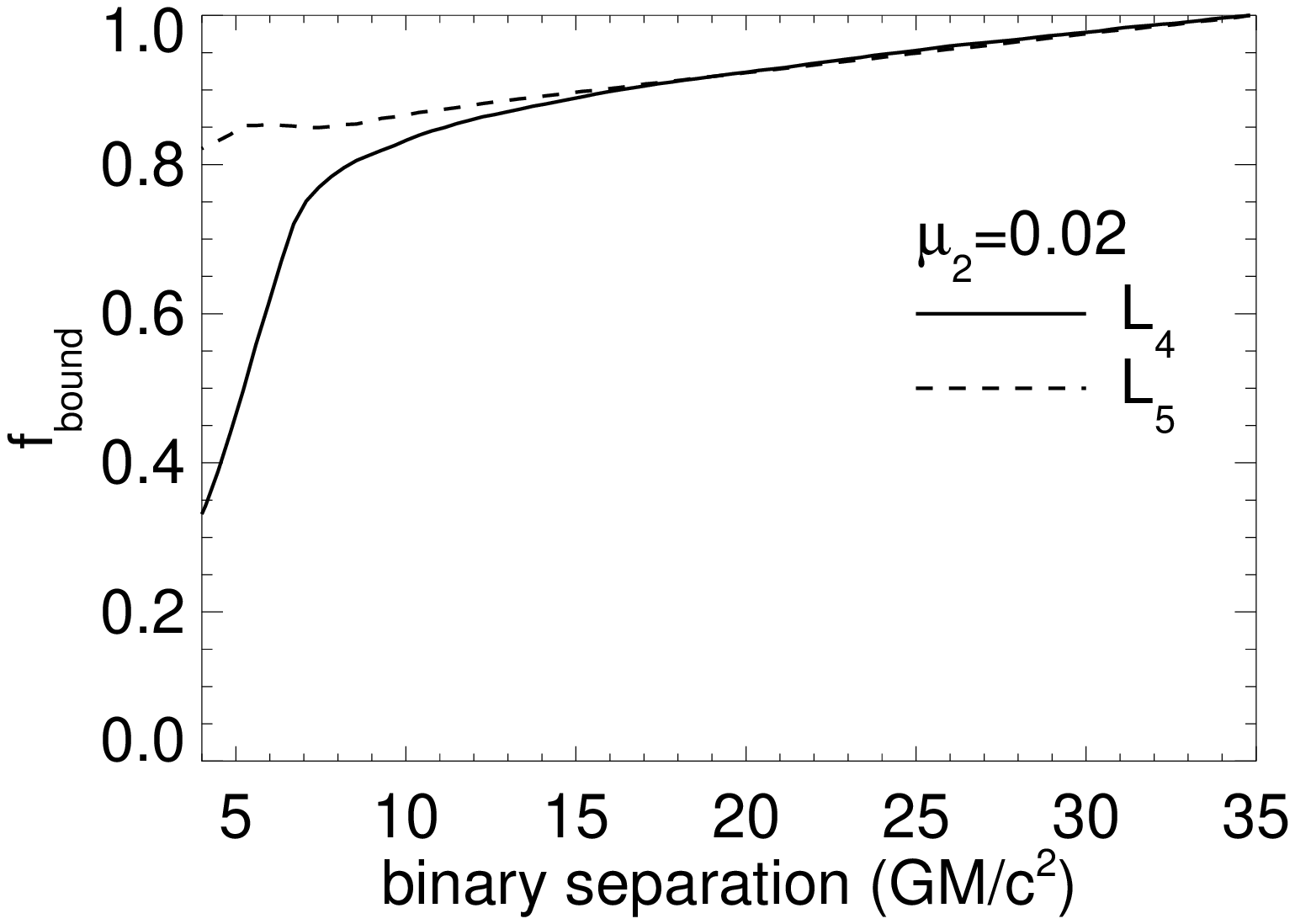}}
\scalebox{0.37}{\includegraphics*[160,425][540,700]{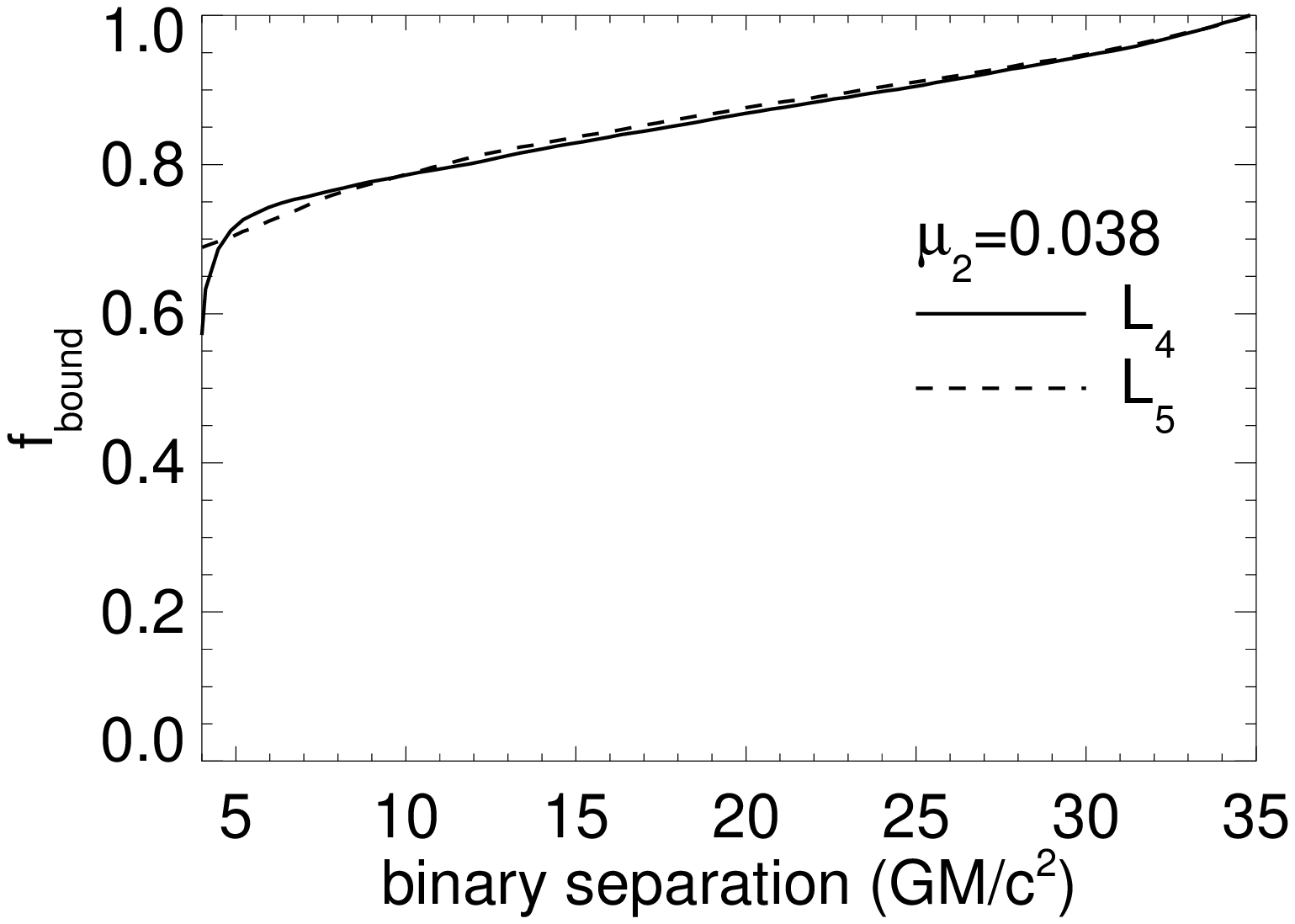}}\\
\scalebox{0.37}{\includegraphics*[55,360][540,700]{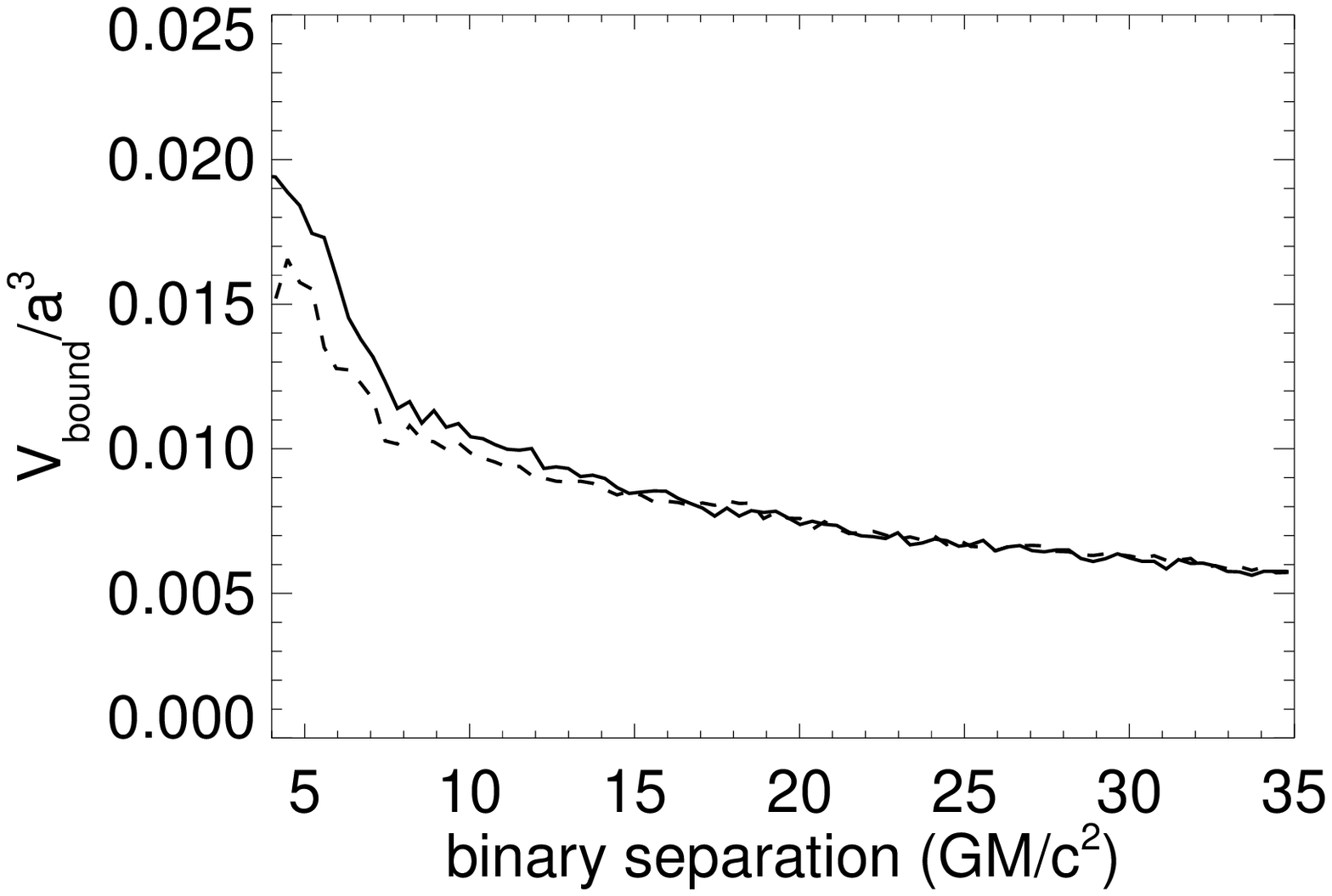}}
\scalebox{0.37}{\includegraphics*[160,360][540,700]{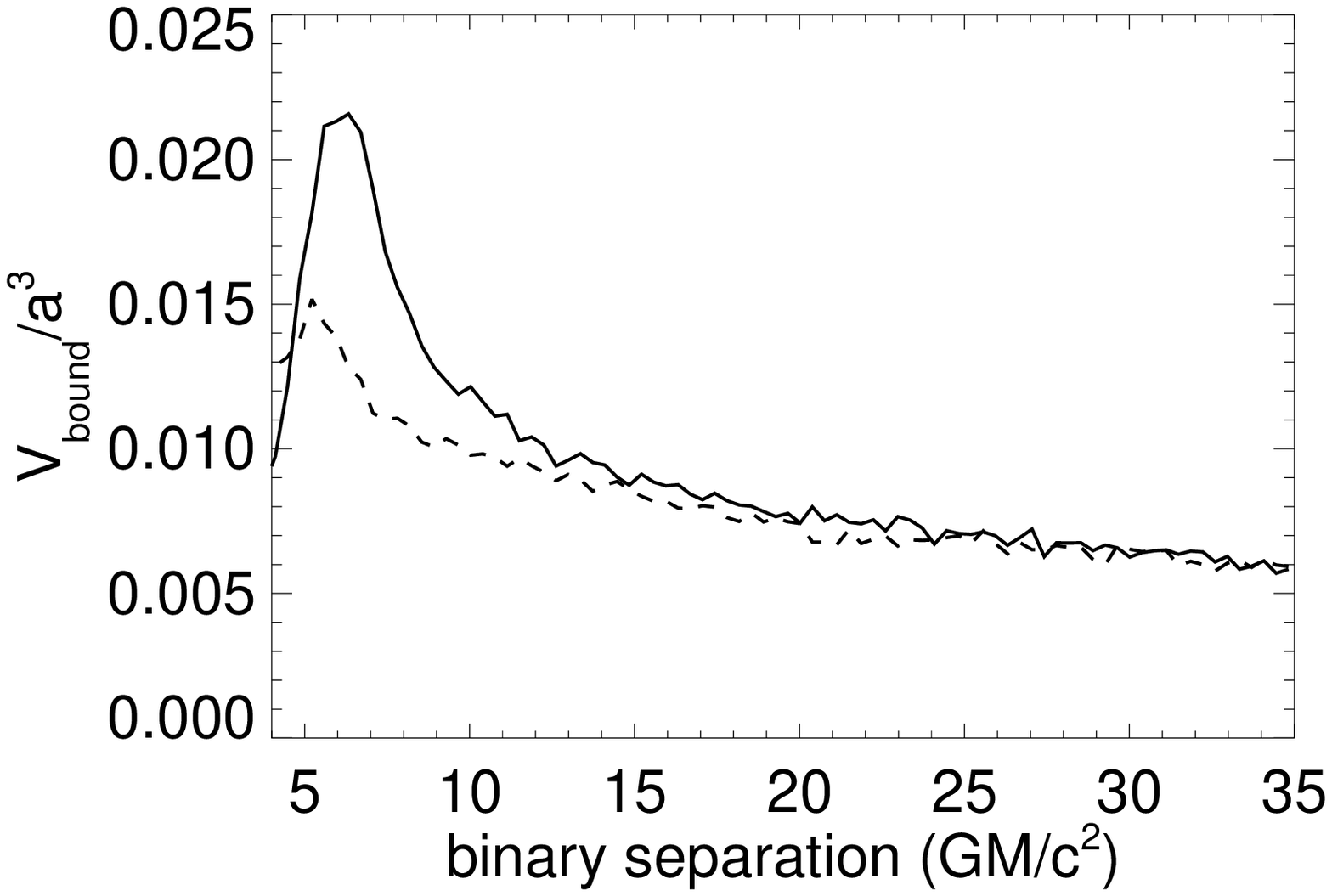}}
\scalebox{0.37}{\includegraphics*[160,360][540,700]{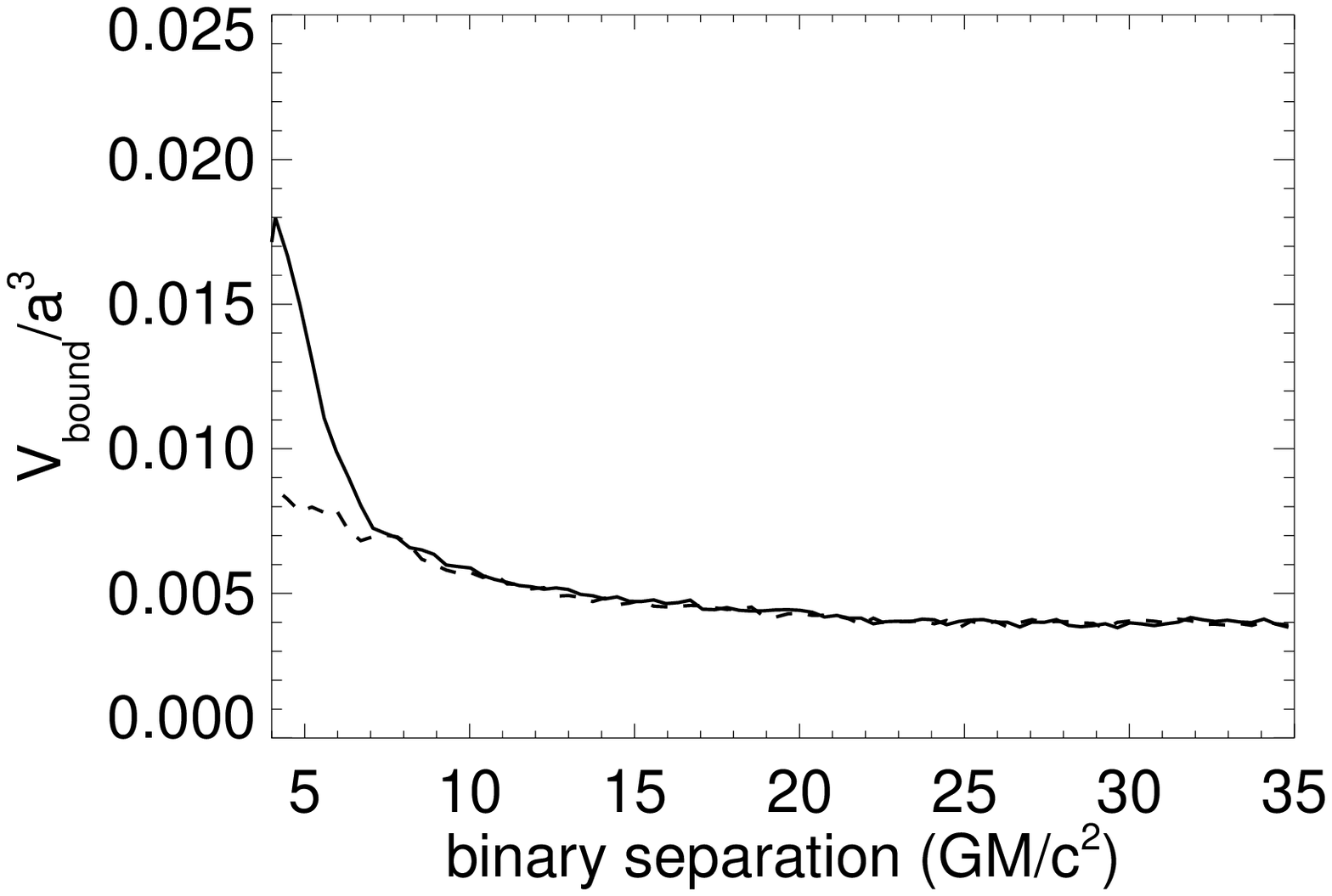}}
\caption{\label{N45_mu} \textit{Upper:} The fraction $f_{\rm bound}$
  of test particles
  that remain bound in librating orbits around $L_4$ ({\it solid
    curves}) and $L_5$ ({\it dashed curves}) for a range of mass
  ratios. The particles are initially distributed in a small cloud
  around each Lagrange point with binary separation $a=40M$. $f_{\rm
    bound}$ is normalized to the number of bound particles when
  $a=35M$. \textit{Lower:} The volume of space occupied by bound test
  particles around $L_4$ ({\it solid curves}) and $L_5$ ({\it dashed
  curves}), normalized by the characteristic volume of the system,
  $a^3$.} 
\end{center}
\end{figure*}

When studying inertial drag forces such as those described above,
\citet{murray:94} found that
the $L_4$ point was more stable than $L_5$. Since our
problem is the mirror-image of that one, it is not surprising then
that we find greater stability around $L_5$ when including GW
evolution. As shown in the upper panel of Figure \ref{dphi} for
$\mu_2=0.038$, a test
particle initially very close to $L_4$ will eventually get ejected
from the equilibrium configuration as the binary approaches merger. On
the other hand, test particles librating around $L_5$ appear to become
{\it more} stable, asymptotically approaching the exact equilibrium
position. As mentioned in \citet{erdi:09}, for binaries with $\mu_2$
slightly above the critical value for linear stability ($0.0385
\lesssim \mu_2 \lesssim 0.0401$), there still
exist bound orbits that librate around $L_4$ and $L_5$ with moderate
amplitude. We find that for these systems, the inclusion of radiation
reaction increases the stability of $L_5$ by damping the libration
amplitude. On the other hand, the libration amplitude of particles
around $L_4$ grows rapidly towards the end of inspiral, as in the
linear case in Figure \ref{dphi}. In contrast, \citet{seto:10} find no
difference between $L_4$ and $L_5$, although they only consider a few
test particles around each Lagrange point.

To further investigate this asymmetry in stability, we carried out
orbital integrations for large numbers of test particles distributed
in clouds around each Lagrange point for masses $\mu_2 = 0.01$,
$0.02$, and $0.038$. This allowed us to probe the regions relatively
far from the equilibrium points, where standard linear stability
analysis breaks down. We used two methods for estimating the
large-scale stability of orbits around each Lagrange point. The first
was a simple number count: the fraction of particles that remain
bound around the equilibrium point. In this context, a particle is
considered bound if its instantaneous Jacobi constant $C_{\rm J}$ is less
than $C_{\rm J0}(L_3)$ for the present orbital separation, corresponding
to the tadpole-shaped regions around $L_4$ and $L_5$ in Figure
\ref{cj_cont}. Additionally, any
particle that escapes with $r>10a$ or approaches closer than $3\mu_1$
from the primary or $3\mu_2$ from the secondary is removed from the
system. The second method of measuring stability was to calculate the
total volume filled by bound particles around each Lagrange point. A
larger volume corresponds to lower stability, as the particles begin
to escape the bound libration region.

The fraction of bound particles $f_{\rm bound}$ is plotted in the top panels
of Figure \ref{N45_mu}, normalized to the bound number at binary
separation of $a=35M$. The particles begin in a small cloud of radius
$\delta R$ around each Lagrange point with random velocities,
isotropic with $\langle v \rangle = v_{\rm orb}(\delta R/a)^{1/2}$ in
the corotating 
frame. We select $\delta R$ to be large enough so that the cloud of
particles roughly fills the phase-space volume of stable orbits around $L_4$
and $L_5$. In practice, this typically corresponds to half of the
particles getting ejected on roughly a dynamical time after the start
of the simulation (hence $f_{\rm bound}$ is defined relative to
$a=35M$, not $a=40M$). After a stable cloud is formed around each Lagrange
point, marginally bound particles are slowly ejected from the system
(or accreted onto one of the BHs) as the binary system evolves
adiabatically. From Figure \ref{dphi}, we might expect more particles
to get ejected from $L_4$, which appears to be less stable, at least
at linear order. However, as evident in Figure \ref{N45_mu}, $L_4$
retains more bound particles when $\mu_2=0.01$, $L_5$ appears more
stable for $\mu_2=0.02$, and they are nearly identical for
$\mu_2=0.038$. 

It is not entirely clear why this peculiar behavior
occurs, with no apparent trend in global stability as a function of
mass ratio, nor why one Lagrange point or the other should be more
stable for a given mass ratio. As described in
\citet{murray:99}, test particles near the stable Lagrange
points move along epicycles with two distinct frequencies. For certain
values of $\mu_2$, these libration frequencies have integer
commensurabilities, which naturally leads to strong resonant
interactions \citep{deprit:67}. For example, when $\mu_2=0.0243$, both
$L_4$ and $L_5$
are stable under standard linear analysis, yet the two epicyclic 
frequencies have a ratio of $2:1$, leading to resonant excitations and
ejection of all test particles after a few dynamical
times \citep{erdi:09,seto:10}. \citet{murray:94} shows that these characteristic
frequencies change in the presence of inertial drag forces, which we
have shown above to be directly analogous to GW evolution. Thus we
suspect that the different behavior seen in $f_{\rm bound}(t)$ for
different Lagrange points is due to a combination of resonant
interactions and asymmetric shifts in the libration frequencies due to
the binary evolution via GW losses.

In the bottom panels of Figure \ref{N45_mu} we plot the total volume
$V_{\rm bound}$ occupied by bound particles, normalized to the overall scale of the
binary $a^3$. In the limit of a very small $\delta R/a$, where the
test particles are well within the linear regime, the cloud around
each Lagrange point will actually expand adiabatically relative to the
binary separation, although shrinking on an absolute
scale. \citet{fleming:00} show this evolution to scale as $a^{9/4}$:
\begin{equation}
\left(\frac{V_f}{V_i}\right) = \left(\frac{a_f}{a_i}\right)^{9/4},
\end{equation}
where $V_i$ and $V_f$ are the initial and final volumes of test
particle clouds. We have confirmed this result numerically with 
simulations starting with $\delta R \ll a$. 
On the other hand, if $\delta R/a$ is large enough such that
the stable region of phase space is completely filled, the
normalized volume $V/a^3$ should be constant, since the entire problem is
scale-invariant. Any adiabatic expansion will not increase the volume
of bound orbits, but rather decrease the total number of particles in
these regions, as seen in the upper panels of Figure \ref{N45_mu}. 

We actually find a gradual increase in $V_{\rm bound}/a^3$ as the
binary shrinks, yet still much
slower than the \citet{fleming:00} result of $(V/a^3)\sim a^{-3/4}$, suggesting that the bound
region in phase space is in fact filled. The increase in $V_{\rm
  bound}/a^3$ during the adiabatic inspiral phase ($a\gtrsim 10M$) is
likely due to a combination of effects. First, we expect a net 
increase in global stability caused by the GW losses, analogous to the
effects of inertial drag described above. Secondly, there is also some
inherent error in measuring the volume of bound orbits, based on our
crude metric of the test particles' instantaneous Jacobi
constants. Some fraction of particles that have moved outside the
stable region may still appear bound for a few dynamical times
before getting ejected or accreted, and thus artificially contribute
to $V_{\rm bound}$. At smaller $a$, where the libration 
time becomes comparable to the inspiral time, the system no longer
evolves adiabatically, and a cloud of particles can get ``frozen'' at
roughly constant volume, leading to the rapid increase in $V_{\rm
  bound}/a^3$ for $a\lesssim 10M$. Interestingly, we do see a
consistent trend of larger $V_{\rm bound}$ around $L_4$ than $L_5$ for
all values of $\mu_2$, as
suggested by Figure \ref{dphi}. However, this may be less a measure of
local stability than a consequence of the fact that $L_4$ moves {\it
  toward} the secondary, while $L_5$ moves away from it. Thus the two
regions begin to sample different potentials, and the gravitational
force of the secondary acts to disrupt the cloud around $L_4$ more
strongly than that around $L_5$, in turn leading to a larger $V_{\rm
  bound}$. 

We should add as a word of caution, that many of these numerical
stability results may be altered somewhat by the inclusion of
additional PN terms to the equations of motion \citep{seto:10}. In general
relativity, unlike the case of Newtonian gravity, test particles moving in
the Kerr metric have three distinct epicyclic frequencies. These
additional frequencies may lead to more complicated resonant
interactions and thus change the behavior of $f_{\rm bound}(a)$ and
$V_{\rm bound}(a)$, at least at a quantitative level. Yet our
intuition suggests that the {\it qualitative} behavior shown in Figure
\ref{N45_mu} should remain unchanged. Future work will investigate
this question in greater detail.

%--------------------------------------------------------
\section{ASTROPHYSICAL APPLICATIONS}
\label{applications}
%--------------------------------------------------------

\subsection{Formation Mechanisms}
In the previous Sections, we showed analytically and numerically how
the location and stability of the $L_4$ and $L_5$ Lagrange points
evolve in the presence of radiation reaction. We found that test
particles, once captured into libration orbits around one of the
stable Lagrange points, will in general remain there throughout
the adiabatic inspiral phase, as the entire length scale of the system
steadily shrinks. However, the initial premise was not addressed: will
there even be gas or stars or debris present in the first place? How
might matter get captured into these orbits at large binary
separations of $a \gtrsim 1$ pc? Furthermore, are binary BHs with
$\mu_2 \lesssim 0.04$ astrophysically likely, or even possible?

Even in a system as well-studied as the Sun-Jupiter binary, there is
still no clear consensus on the origin of the Trojan asteroids,
despite the fact that orbital elements are now known for over 4000
individual objects. \citet{rabe:72} proposed that they may be
captured comets, while \citet{shoemaker:89} and \citet{kary:95}
suggested a more local origin as planetesimals in the early solar
system. In most theories, some form of dissipative force is required to capture the
Trojans \citep{yoder:79}, such as nebular gas drag \citep{murray:94,kary:95} or
planetesimal collisions \citep{shoemaker:89}. Additionally, an
increase in Jupiter's mass via accretion of the surrounding disk could
lead to asteroids getting captured
\citep{marzari:98,fleming:00}. Analogous to our BH binary system, the
inward migration of Jupiter may have also affected the capture and
subsequent evolution of the Trojans \citep{yoder:79,fleming:00}.

The characteristic separation at which point a hard BH binary (i.e.,
the orbital potential is dominated by the two BHs, and not by the
surrounding stars) is
formed is comparable to the primary BH's radius of influence
\citep{merritt:06}: 
\begin{equation}\label{a_h}
a_{\rm h} = \frac{q}{(1+q)^2}\frac{R_{\rm infl}}{4},
\end{equation}
where $q=M_2/M_1$ and 
\begin{equation}
R_{\rm infl} = \frac{GM}{\sigma^2} \approx 1.1\, {\rm pc}\,
\left(\frac{M}{10^7 M_\odot}\right)\left(\frac{\sigma}{200\, {\rm
    km}\, {\rm s}^{-1}}\right)^{-2}\, ,
\end{equation}
with $\sigma$ being the one-dimensional velocity dispersion of the
galaxy's bulge. In the limit of small $\mu_2 \approx q\ll 1$, we have
\begin{equation}\label{a_h2}
a_{\rm h} \approx 5\times 10^{-3}\, {\rm pc} 
\left(\frac{\mu_2}{0.02}\right)
\left(\frac{M}{10^7 M_\odot}\right)\left(\frac{\sigma}{200\, 
{\rm km}\, {\rm s}^{-1}}\right)^{-2}\, ,
\end{equation}
corresponding to $a_{\rm h} \sim 10^4 M$ in geometrized units.
Within this region, there are typically thousands of stars
\citep{hopman:06}, and in active galactic nuclei (AGN), a great deal
of gas and dust 
as well. It is also roughly the separation at which point the binary
will merge within a Hubble time due to GW losses alone
\citep{peters:64}, and thus equations (\ref{eoma})-(\ref{eomd}) are
appropriate.  

For AGN with massive accretion disks, it is possible for significant
star formation to occur in the region where the disk becomes
self-gravitating \citep{levin:03}. For standard thin disk models
\cite{shakura:73}, this
self-gravitating radius actually corresponds closely to $a_{\rm
  h}$. Since galaxy mergers are often associated with gas inflow and
quasar activity, it is quite
possible that stars could form in an accretion disk just as a hard
BH binary is forming, thereby trapping the stars around the stable
Lagrange points. A similar possibility, suggested by \citet{goodman:04},
is that supermassive stars with $M\gtrsim 10^4 M_\odot$ could grow out
of main-sequence seeds embedded in the accretion disk, quickly
evolving and then collapsing to intermediate-mass black holes
(IMBHs). This scenario is particularly attractive, as it naturally
provides a mechanism for making a binary BH with $\mu_2 \lesssim 0.01$
on a circular orbit, surrounded by ample debris that could easily get
trapped into resonant orbits. 

Another mechanism that could explain both the low-mass secondary and
also provide a supply of stars would be the tidal disruption of a
globular cluster by the primary BH. If the globular cluster hosts an
IMBH \citep{miller:02}, it could form a hard binary with
the primary, and 
the surrounding stars could get captured into stable orbits around
$L_4$ and $L_5$. At this point, it would be impossible to carry out
any quantitative analysis of the relative rates and likelihoods of
these different scenarios, considering the remarkable lack of
observational evidence for even a single SMBH binary with orbit smaller than
$a_{\rm h}$. Rather, we might realistically hope to discover such a system
with help from one of the EM signatures described below.

\subsection{Tidal Disruptions}

Tidal disruptions of solar-type stars by SMBHs are expected to occur
at a rate of $\sim 10^{-4}$ yr$^{-1}$ per MW-type galaxy
\citep{wang:04}, and are excellent tools for detecting
otherwise-quiescent nuclei at cosmological distances
\citep{lidskii:79,rees:88}. Typical
luminosities are comparable to the Eddington limit, and the timescale
for the rise and subsequent decay of the light curve is on the order
of weeks to months \citep{ulmer:99,halpern:04}, which is ideal for detection by wide-field
time-domain surveys such as Pan-STARRS\footnote{\tt
  http://pan-starrs.ifa.hawaii.edu} or LSST\footnote{\tt
  http://www.lsst.org}
\citep{gezari:09}. Tidal disruptions have recently been proposed as
tools for detecting BH mergers or recoiling merger remnants, in some
cases long after the merger occurs \citep{komossa:08, chen:09,
stone:10}.

\begin{figure}
\begin{center} 
\scalebox{0.45}{\includegraphics{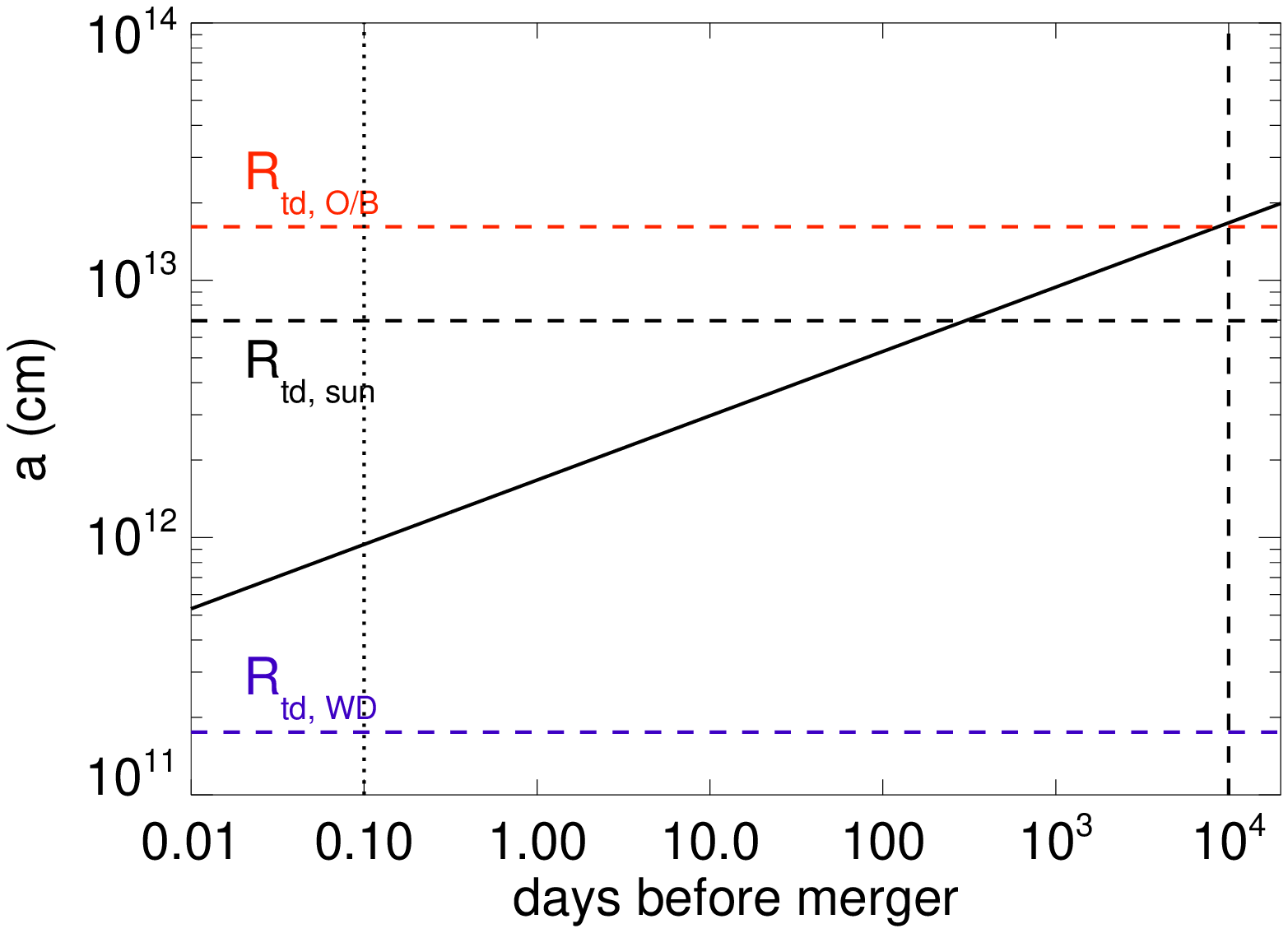}}\\
\scalebox{0.45}{\includegraphics{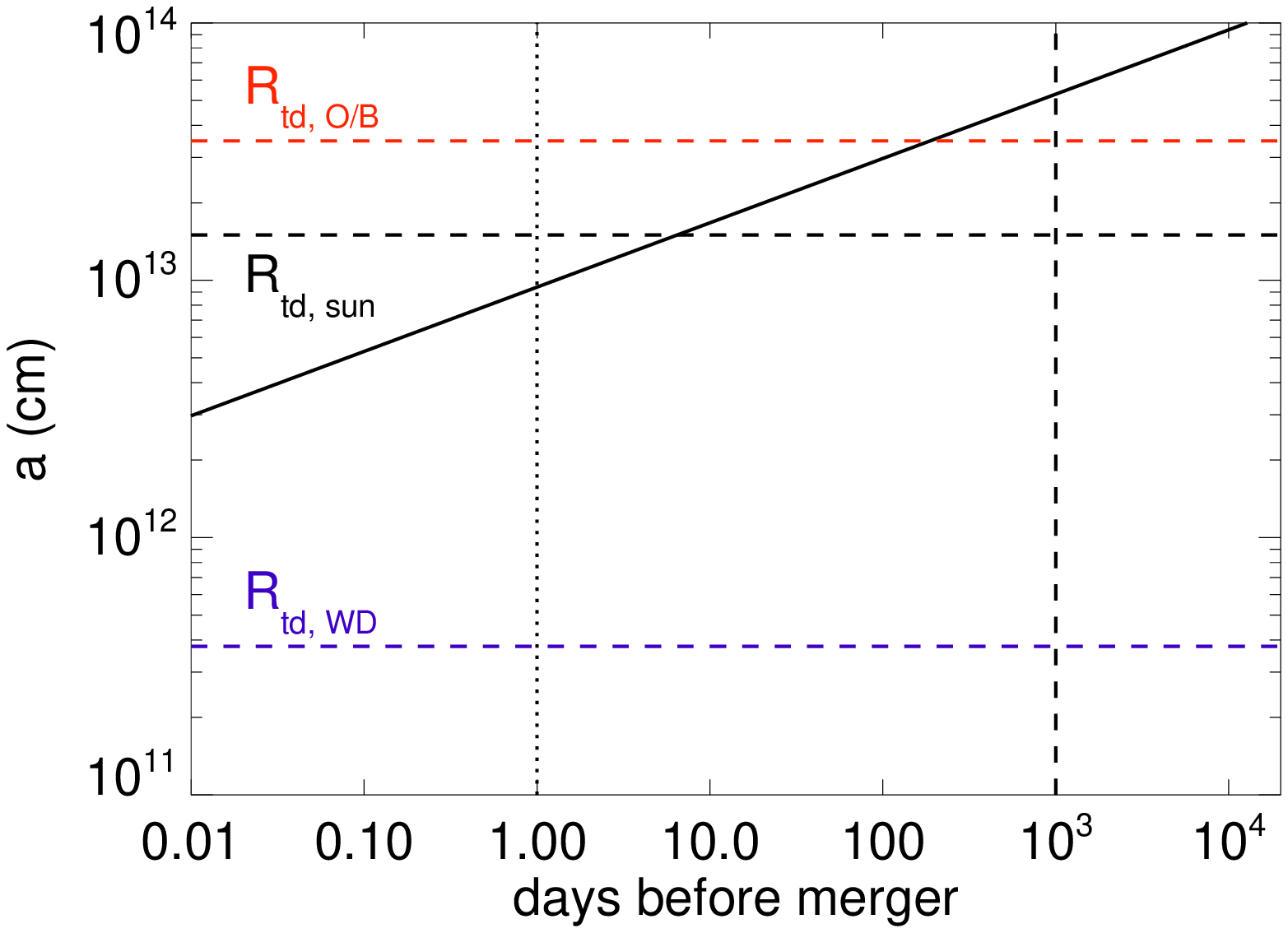}}
\caption{\label{T_td} \textit{Upper:} Tidal disruption radius for a
  star orbiting a $10^6 M_\odot$ BH, as a function of time before merger. The
  secondary BH has mass $10^4 M_\odot$. 
  \textit{Lower:} Tidal disruption radius for a
  star orbiting a $10^7 M_\odot$ BH, as a function of time before merger. The
  secondary BH has mass $10^5 M_\odot$. From top to bottom, the
  horizontal lines correspond to the tidal disruption radius of a main
  sequence O/B star, a solar-type star, and a 0.7 $M_\odot$ white
  dwarf. The vertical dotted line corresponds to the point at which
  the secondary BH reaches the ISCO and the vertical dashed lines the
  point at which the binary enters the LIGO band ($f_{\rm GW} \approx 1$ mHz).}
\end{center}
\end{figure}

In our scenario, the main-sequence stars are captured into resonant
motion around $L_4$ or $L_5$ at large separation ($a \sim a_{\rm h}$)
and then move in closer to the primary as the binary separation slowly
shrinks. A star of mass $M_\ast$ and radius $R_\ast$ will get
disrupted when the tidal forces from the BH
are comparable to the star's own self-gravity, which occurs at a
distance $R_{\rm td}$ from the primary:
\begin{equation}\label{R_td}
R_{\rm td} \approx R_\ast \left(\frac{M_1}{M_\ast}\right)^{1/3}.
\end{equation}
In our numerical tests of Section \ref{numerical_tests}, we found that
during inspiral, some fraction of test particles get ejected from
$L_4$ or $L_5$ and disrupted by $M_2$, the the majority remain bound
until the final merger stage. For stars that remain bound to $L_4$ or
$L_5$, the gravitational tidal forces are dominated by the primary,
and thus the relevant mass in equation (\ref{R_td}) is $M_1$. 

In Figure \ref{T_td} we plot the orbital separation of a BH binary as
a function of time before merger ({\it black curves}). The dashed
horizontal lines correspond to the points at which (top-to-bottom) an
O/B star, a solar-type star, or a white dwarf would get tidally
disrupted by the primary BH of mass $10^6 M_\odot$ ({\it top}) or $10^7
M_\odot$ ({\it bottom}), and secondary mass $\mu_2=0.01$. The vertical
dotted lines denote where the
binary separation reaches the ISCO and the BHs merge, and the vertical
dashed lines correspond to the point where the
binary enters the LISA GW frequency band. For $M_1=10^6 M_\odot$, a
solar-type star would get disrupted roughly a year before
merger. At that point, LISA would already be able to provide an error-box of
$\sim 100$ square degrees for the source location on the sky
\citep{lang:08}, which
could be easily monitored for luminous tidal disruption flares out to
redshifts $z \sim 0.2$ or so \citep{gezari:09}. In all cases with $M_1 \gtrsim 10^5
M_\odot$, white dwarfs would not get disrupted, but rather plunge
directly into the BH. 

\begin{table}
\caption{ \label{table_end}
Fraction of test particles that, during inspiral, are either
 ejected from the system or captured by one of the BHs.}
\begin{center}
\begin{tabular}{lccc}
 & $\mu_2=0.01$ & $\mu_2=0.02$ & $\mu_2=0.038$ \\
\hline
$f(L_4\to \infty)$ & 0.13 & 0.11 & 0.13 \\
\vspace{0.1cm}
$f(L_5\to \infty)$ & 0.20 & 0.12 & 0.21 \\
$f(L_4\to M_1)$ & 0.16 & 0.33 & 0.19 \\
\vspace{0.1cm}
$f(L_5\to M_1)$ & 0.15 & 0.05 & 0.10 \\
$f(L_4\to M_2)$ & 0.14 & 0.27 & 0.18 \\
$f(L_5\to M_2)$ & 0.21 & 0.11 & 0.20
\end{tabular}
\end{center}
\end{table}

In addition to the stars that get disrupted at late times, we saw in
the simulations of Section \ref{numerical_tests} that there are
also some particles that get captured by either $M_1$ or $M_2$ during
the inspiral phase, thus leading to tidal disruption, or get ejected
entirely from the system. Table \ref{table_end} shows the fraction of
test particles that meet one of these various fates during inspiral,
i.e., when the binary separation is $5M \le a(t) \le 35M$. Of course,
if equation (\ref{R_td}) is met, then even stars on stable orbits are
disrupted. As in Figure \ref{N45_mu}, there are no clear trends with
varying $\mu_2$, but we do confirm that particles around $L_5$ are
generally more likely to get ejected from the system, while particles
around $L_4$ are more likely to get captured or disrupted by the BHs.

\subsection{Ultra-Hyper-Velocity Stars}

Of the stars that {\it do} get ejected from the binary system during
inspiral, we find a wide range in their velocities at infinity. To first
order, one would expect the ejection velocity to be comparable to the
orbital velocity. However, particles ejected at late times when the
orbital velocity is highest are also coming from the deepest
potentials, so are slowed more as they escape from the system. In
practice, we find that net effect is a slight increase in the average escape
velocity at late times, but somewhat flatter than the orbital velocity
scaling of $a^{-1/2}$. This is likely due to the fact that the most
efficient three-body slingshot interactions require the test particle
to come within a small fraction of $a$ from the secondary, and when
$a$ becomes smaller, these close approaches more likely lead to
capture or disruption than ejection. 

From equation (\ref{R_td}), we see that a solar-type star would get
tidally disrupted by a $10^6 M_\odot$ BH at a distance of $\sim 50M$,
far outside the ISCO. For $M_1=10^7 M_\odot$, however, a star could
survive to within $10M$ of the primary. From that separation, we find
that test particles are ejected with mean velocities of $0.1c$,
roughly 50 times greater than the radial velocities of hyper-velocity
stars (HVSs) observed in the outer regions of the Milky Way
\citep{brown:09}. Of course, any star moving that fast would be
observable for a much shorter time; only a few million years after the
BH merger. At the same time, this allows younger, more massive stars
to survive out to the edge of the galaxy, making them even easier to
identify against the field stars. It is conceivable that the BH at the
center of our galaxy
may have experienced a minor merger with $\mu_2 \sim 0.01$ in the past
10 Myr, an event which may be convincingly inferred by the detection of
such an ultra-HVS at a distance of $\sim 50-100$ kpc. 

\subsection{Compressed Stellar Systems}

In the event that many stars are initially trapped around $L_4$ or
$L_5$, the gravitational interactions between the test particles may
become important, an effect that we have neglected in this
work. Especially for larger systems with $M \gtrsim 10^8 M_\odot$,
one could even imagine forming globular clusters around the Lagrange
points, although from equation (\ref{a_h2}) it is clear they would
necessarily be quite compact. When the core of a normal globular
cluster grows too dense, it can undergo a collapse via gravitational
instability, at the same time expanding the orbits of the outer stars,
due to conservation of energy. In our BH binary system, the adiabatic
compression of orbits may trigger core collapse of the trapped
cluster, while simultaneously preventing any outer stars from
escaping, thereby enhancing the stellar density further. If the
density becomes high enough, stellar collisions will become common,
and the system could act as a breeding ground for supermassive stars
and IMBHs \citep{portegies:99}.

\subsection{Compact Objects}

Many of the formation scenarios considered above naturally lead to
high-mass stars forming in the dense gaseous environments around the
SMBH binary. These stars could then evolve and collapse into neutron
stars or stellar-mass BHs on timescales short compared to the inspiral
time. Unlike main sequence stars that would get tidally disrupted
well before the merger, compact objects would survive unperturbed
until they either get ejected or captured whole by one of the BHs. In
that case, they would also contribute to the GW signal produced by the
binary. For a system with $M_1=10^6 M_\odot$, $M_2=10^4 M_\odot$, and
$m_0= 10 M_\odot$, the stellar-mass BH would contribute an additional
$0.1\%$ to the waveform amplitude. At a distance of 1 Gpc, this small
perturbation would still be observable with LISA. 

However, if the compact object is orbiting exactly at the $L_4$ or
$L_5$ with no libration, it would contribute a pure sinusoid component
to the waveform with identical frequency as the dominant wave, only with a
phase shift of $\pm 120^\circ$. Thus it would be
indistinguishable from a normal binary on a circular orbit with
slightly different masses and phase. On the other hand, if the compact object is
librating with typical amplitude $\Delta \phi \approx 30^\circ$, as we
might expect in the general case, the waveform will contain a distinct
signal at the libration frequency (in this example, $\omega_{\rm lib}
\approx 0.27 \omega_{\rm orb}$) which may still be detectable. More
likely, this extra signal would be confused with a spin-orbit or
spin-spin precession frequency, effectively degrading the overall
confidence in the binary parameter estimation. In some cases, the
stellar-mass BH might get ejected during inspiral, but with small
enough velocity that it doesn't escape the system, but rather remains
bound on a larger, eccentric orbit even after the two massive BHs
merge, ultimately merging much later as an extreme mass-ratio inspiral
(EMRI) source.

As mentioned above, the secondary BH and the ``test particle'' can
actually have comparable masses and still maintain stability of the Lagrange
points, as long as $\mu_2 \lesssim 0.0385$. Thus a stable three-body
BH system could exist with $M_1 \approx 10^6 M_\odot$ and $M_2 \approx
M_3 \approx 10^4 M_\odot$, in which case the orbital libration could
be seen very clearly in a LISA waveform. The probability of such a
system existing is perhaps not as small as one might think, especially
if both IMBHs form out of a massive accretion disk, not unlike
Saturn's moons Janus and Epimetheus. Indeed, only 15 years ago, the
possibility of ``Hot Jupiters'' with $\mu_2 \gtrsim 0.001$, orbiting
very close to their host stars was considered quite remote, and now we
know of hundreds of such systems. The same may very well be true of
SMBHs and their surrounding gas disks. 

\subsection{Diffuse Gas}

In the circumbinary accretion disk model, ample gas and dust will
surround the BH binary, of which some fraction could likely get
trapped into orbits around $L_4$ and $L_5$. If cold enough, this gas
could get compressed by the evolving binary potential and form stars
or even clusters of stars, as discussed above. More likely, the gas
will be rather hot, with thermal velocities comparable to the
libration velocities of test particles filling the stable volume 
around $L_4$ and $L_5$. One very rough estimate of this thermal motion
may come from inspection of equation (\ref{C_J}) and Figure
\ref{cj_cont}. Trapped particles will have $C_{\rm J0}(L_4)\lesssim C_{\rm J}
\lesssim C_{\rm J0}(L_3)$, so typical velocities in the rotating frame of
\begin{equation}\label{v2lib}
v^2 \approx C_{\rm J}(L_3,v=0)-C_{\rm J}(L_4,v=0)\approx 2\mu_2 a^{-1}
\end{equation}
\citep{murray:99}.
Combining with equation (\ref{a_h2}) for the separation at the time of
binary formation, we get an initial thermal velocity of 
\begin{equation}\label{v_therm}
v_{\rm therm} \approx 600\, {\rm km}\, {\rm s}^{-1}
\left(\frac{\sigma}{200\,  {\rm km}\, {\rm s}^{-1}}\right),
\end{equation}
or roughly $4\times 10^7$ K. Somewhat remarkably, $v_{\rm therm}$ is
independent of $\mu_2$ or $M_1$ (except indirectly
through the $M-\sigma$ relationship). 

We can estimate an upper limit on the density of this gas $\rho_0$ by
requiring it to be stable to gravitational collapse. In other words,
the total mass should not exceed the Jeans mass:
\begin{equation}
0.005\, \rho_0 a_{\rm h}^3 \le \left(\frac{5k_B T}{Gm_p}\right)^{3/2}
  \left(\frac{3}{4\pi \rho_0}\right)^{1/2},
\end{equation}
with $k_B$ the Boltzmann constant and $m_p$ the proton mass. The
leading factor of $0.005$ comes from the volume of the stable region,
estimated from Figure \ref{N45_mu}. Taking $m_p v_{\rm therm}^2 = 3
k_B T$, we get 
\begin{equation}
\rho_0 \le 35 \left(\frac{v_{\rm therm}^2}{G a_{\rm h}^2}\right),
\end{equation}
corresponding to a total mass in gas of
\begin{equation}
M_{\rm gas} \approx 0.005 \rho_0 a_{\rm h}^3 \le
7\times 10^4 M_\odot
\left(\frac{\mu_2}{0.02}\right)
\left(\frac{M}{10^7 M_\odot}\right)\, ,
\end{equation}
which is far more than enough material to fuel a bright EM
counterpart.  

Even if the density of the trapped gas is orders of magnitude smaller,
it should still be sufficient to produce observable emission lines if
irradiated by an AGN or quasar powered by accretion onto one of the
BHs. For a cloud of gas filling the stable region around each Lagrange
point, the covering fraction would be on the order of a few percent. Since the
thermal velocity is roughly a factor of 10 smaller than the orbital
velocity, the emission lines would be somewhat narrower than a
classical broad line originating from the same radius. The coherent
shape and motion of the cloud would also give rise to a large velocity
offset relative to the host galaxy or narrow emission
line clouds residing at a greater distance from the central
engine. This offset in wavelength would vary periodically with 
the binary orbit ($T_{\rm orb}$ at our fiducial value of $a_h$ would
be on the order of 10 years), giving strong evidence for the existence of a BH
binary. Such an effect could be detected by large spectroscopic
surveys like the Sloan Digital Sky Survey, a technique already in
frequent use by
observers searching for binary quasars or recoiling black holes
\citep{bonning:07,komossa:08b,boroson:09}

%--------------------------------------------------------
\section{DISCUSSION}
\label{discussion}

We have analyzed the circular restricted three-body problem
including leading-order terms for gravitational wave losses, in
particular studying the behavior of test particle orbits near the
stable Lagrange points $L_4$ and $L_5$. We find that these orbit
remain stable throughout most of the adiabatic inspiral phase, and
their locations shift relative to the Newtonian limit as the binary
enters the late phases of inspiral and merger. Analogous to the case
of inertial drag forces in the planetary problem, the non-conservative
GW loss terms lead to an asymmetry between $L_4$ and $L_5$: test
particles around $L_4$ move closer to the secondary BH, while those
initially around $L_5$ move away from it. While there appear to be
complicating resonances that affect the relative stability of the two
regions differently for different mass ratios, a general trend is that
the inclusion of GW losses increases the stability of $L_5$ while
decreasing the stability of $L_4$. Particles initially around $L_4$
are more likely than those around $L_5$ to get captured or disrupted
by $M_2$, while particles near $L_5$ are more likely to get ejected
from the system entirely. However, in both cases, the vast majority of
particles get neither ejected nor captured until the binary separation
is on the order $a\lesssim 5M$, at which point the Newtonian equations
of motion break down entirely and everything plunges into the primary
BH.  

The stability of the $L_4$ and $L_5$ Lagrange points throughout most
of the inspiral provides a natural mechanism for shepharding material
from relatively large radii down to the central BH, in turn producing
a luminous EM counterpart to the GW signal from a BH merger. The detailed physics
behind such an EM signal are well beyond the scope of this paper, but
early results from numerical relativity simulations suggest that
bright EM flares should be quite general, provided enough gas is
available \citep{vanmeter:10,bode:10,farris:10}. In addition to the prompt
emission associated with the final merger, we have shown that
hyper-velocity stars with speeds as high as $0.1c$ might be
observable, flying away from the galactic center for millions of years
after the merger. With sufficient column densities of diffuse gas
trapped around $L_4$ and $L_5$, a precursor signal could take the form
of highly red/blue-shifted emission lines, narrower than typical broad
lines, with the line offset varying periodically with the BH
orbit. 

For the majority of the results in this paper, we require that the
secondary mass be relatively small: $\mu_2 \le 0.0385$ to guarantee
linear stability of $L_4$ and $L_5$. It is unclear how common BH binaries
with such mass ratios might be, and whether they are more likely to
form via galactic mergers or {\it in situ}, via the collapse of
supermassive stars, accretion from a surrounding disk, or capture from
a tidally stripped globular cluster. However, there are other
resonances in the restricted three-body problem that may lead to
qualitatively similar behavior: the stable transport of matter from
large radii down to the ISCO at the time of merger. One example from
the solar system is the Hilda family of asteroids that occupy stable
eccentric orbits with periods $2/3$ that of Jupiter. This ensures that
they avoid close encounters with Jupiter at apocenter, where they
spend most of their time. In the corotating frame, a snapshot of their
positions will appear to fill out a triangle with vertices aligned
with $L_3$, $L_4$, and $L_5$ \citep{broz:08}. \citet{gueron:06}
showed that this resonance is stable even in the extreme relativistic
limit, but still with small mass ratios ($\mu_2 \approx 0.001$).
Our own preliminary results
suggest that this $2:3$ resonance is linearly stable for mass ratios
at least as large as $\mu_2 \approx 0.1$, but we leave a comprehensive
study of the problem, including GW losses, to a future paper. 

In addition to all the Newtonian resonance behavior present in the
three-body problem, the two BHs may also get locked into spin-orbital
resonances with each other during inspiral
\citep{schnittman:04}. Coupled with the
non-degenerate orbital frequencies of test particles in a Kerr
background, the inclusion of spinning BHs would introduce many new
degrees of freedom that may affect the stability of $L_4$ and
$L_5$. While we do not think additional PN terms in the equations of
motion will significantly change our results on a qualitative level,
future work that addresses these questions in more detail is certainly
warranted.  

\vspace{0.35cm}\noindent The author would like to thank Doug Hamilton, Matthew
Holman, Scott Hughes, Julian Krolik, David Merritt, and Cole Miller
for helpful discussions and comments. This work was supported by the
Chandra Postdoctoral Fellowship Program.

%--------------------------------------------------------


\begin{thebibliography}{99}
%--------------------------------------------------------

\bibitem[Armitage \& Natarajan(2002)]{armitage:02} Armitage, P.\ J.,
  \& Natarajan, P. 2002, ApJ 567, L9.

\bibitem[Artymowicz \& Lubow(1994)]{artymowicz:94} Artymowicz, P., \&
  Lubow, S.\ H. 1994, ApJ 421, 651.

\bibitem[Artymowicz \& Lubow(1996)]{artymowicz:96} Artymowicz, P., \&
  Lubow, S.\ H. 1996, ApJ 467, L77.

\bibitem[Bloom et al.(2009)]{bloom:09} Bloom, J.\ S. 2009, White paper
  for Astro2010 Decadal Survey, [arXiv:0902.1527].

\bibitem[Bode et al.(2010)]{bode:10} Bode, T., Haas, R., Bogdanovic,
  T., Laguna, P., \& Shoemaker, D. 2010, ApJ 715, 1117.

\bibitem[Bonning et al.(2007)]{bonning:07} Bonning, E.\ W., Shields,
  G.\ A., \& Salviander, S. 2007, ApJ 666, L13.

\bibitem[Boroson \& Lauer(2009)]{boroson:09} Boroson, T.\ A., \&
  Lauer, T.\ R. 2009, Nature 458, 53.

\bibitem[Brown et al.(2009)]{brown:09} Brown, W.\ R., Geller, M.\ J.,
  \& Kenyon, S.\ J. 2009, ApJ 690, 1639.

\bibitem[Bro\u{z} \& Vokrouhlick\'{y}(2008)]{broz:08} Bro\u{z}, M., \&
  Vokrouhlick\'{y}, D. 2008, MNRAS 390, 715.

\bibitem[Buonanno et al.(2009)]{buonanno:09} Buonanno, A., Iyer, B.\
  R., Ochsner, E., Pan, Y., \& Sathyaprakash, B.\ S. 2009, Phys.\
  Rev.\ D 80, 084043.

\bibitem[Chang et al.(2009)]{chang:09} Chang, P., Strubbe, L.\ E.,
  Menou, K., \& Quataert, E. 2009, [arXiv:0906.0825].

\bibitem[Chen et al.(2009)]{chen:09} Chen, X., Madau, P., Sesana, A.,
  \& Liu, F.\ K. 2009, ApJ 697, L149.

\bibitem[Deprit \& Deprit-Bartholom\'{e}(1967)]{deprit:67} Deprit, A.,
  \& Deprit-Batholom\'{e}, A. 1967, AJ 72, 173.

\bibitem[Erdi et al.(2009)]{erdi:09} Erdi, B., Forgacs-Dajka, E.,
  Nagy, I., \& Rajnai, R. 2009, Celest.\ Mech.\ Dyn.\ Astr.\ 104,
  145. 

\bibitem[Farris et al.(2010)]{farris:10} Farris, B.\ D., Liu, Y.-K.,
  \& Shapiro, S.\ L. 2010, Phys.\ Rev.\ D 81, 084008.

\bibitem[Fleming \& Hamilton(2000)]{fleming:00} Fleming, H.\ J., \&
  Hamilton, D.\ P. 2000, Icarus 148, 479.

\bibitem[Gezari et al.(2009)]{gezari:09} Gezari, S., et al. 2009, ApJ
  698, 1367.

\bibitem[Goodman \& Tan(2004)]{goodman:04} Goodman, J., \& Tan, J.\
  C. 2004, ApJ 608, 108.

\bibitem[Gu\'{e}ron(2006)]{gueron:06} Gu\'{e}ron, E. 2006, Class.\
  Quantum Grav.\ 23, 673.

\bibitem[Halpern et al.(2004)]{halpern:04} Halpern, J.\ P., Gezari,
  S., \& Komossa, S. 2004, ApJ 604, 572.

\bibitem[Hopman \& Alexander(2006)]{hopman:06} Hopman, C., \&
  Alexander, T. 2006, ApJ 645, 1152.

\bibitem[Kary \& Lissauer(1995)]{kary:95} Kary, D.\ M., \& Lissauer,
  J.\ J. 1995, Icarus 177, 1.

\bibitem[Komossa \& Merritt(2008)]{komossa:08} Komossa, S., \&
  Merritt, D. 2008, ApJ 683, L21.

\bibitem[Komossa et al.(2008)]{komossa:08b} Komossa, S., Zhou, H., \&
  Lu, H. 2008, ApJ 678, L81.

\bibitem[Krolik(2010)]{krolik:10} Krolik, J.\ H. 2010, ApJ 709, 774.

\bibitem[Lang \& Hughes(2008)]{lang:08} Lang, R.\ N., \& Hughes, S.\
  A. 2008, ApJ 677, 1184.

\bibitem[Levin \& Beloborodov(2003)]{levin:03} Levin, Y., \&
  Beloborodov, A.\ M. 2003, ApJ 590, L33.

\bibitem[Lidskii \& Ozernoi(1979)]{lidskii:79} Lidskii, V.\ V., \&
  Ozernoi, G.\ M. 1979, AZh.\ Pis'ma 5, 28 (English transl. in Sov.\
  Astron.\ Lett.\ 5, 16).

\bibitem[O'Neill et al.(2009)]{oneill:09} O'Neill, S.\ M., Miller,
  M.\ C., Bogdanovic, T., Reynolds, C.\ S., \& Schnittman, J.\
  D. 2009, ApJ 700, 859.

\bibitem[Marzari \& Scholl(1998)]{marzari:98} Marzari, F., \& Scholl,
  H. 1998, Icarus 131, 41.

\bibitem[Merritt(2006)]{merritt:06} Merritt, D. 2006, ApJ 648, 946.

\bibitem[Miller \& Hamilton(2002)]{miller:02} Miller, M.\ C., \&
  Hamilton, D.\ P. 2002, MNRAS 330, 232.

\bibitem[Milosavljevic \& Phinney(2005)]{milos:05} Milosavljevic, M.,
  \& Phinney, E.\ S. 2005, ApJ 622, L93.

\bibitem[Mosta et al.(2010)]{mosta:10} Mosta, P., Palenzuela,
  C., Rezzolla, L., Lehner, L., Yoshida, S., Pollney, D. 2010, Phys.\
  Rev.\ D 81, 064017.

\bibitem[Murray(1994)]{murray:94} Murray, C.\ D. 1994, Icarus 112,
  465. 

\bibitem[Murray \& Dermott(1999)]{murray:99} Murray, C.\ D., \&
  Dermott, S.\ F. 1999, {\it Solar System Dynamics}, Cambridge
  University Press, Cambridge.

\bibitem[Palenzuela et al.(2009)]{palenzuela:09} Palenzuela, C.,
  Anderson, M., Lehner, L., Liebling, S.\ L., \& Neilsen, D. 2009,
  Phys.\ Rev.\ Lett.\ 103, 081101.

\bibitem[Palenzuela et al.(2010)]{palenzuela:10} Palenzuela, C.,
  Lehner, L., \& Yoshida, S. 2010, Phys.\ Rev.\ D 81, 084007.

\bibitem[Peters(1964)]{peters:64} Peters, P.\ C. 1964, Phys.\ Rev.\
  136, 1224.

\bibitem[Portegies Zwart et al.(1999)]{portegies:99} Portegies Zwart,
  S.\ F., Makino, J., McMillan, S.\ L.\ W., \& Hut, P. 1999, A \& A
  348, 117.

\bibitem[Pringle(1991)]{pringle:91} Pringle, J.\ E. 1991, MNRAS 248,
  754. 

\bibitem[Rabe(1972)]{rabe:72} Rabe, E. 1972, in {\it Motion, Evolution
  of Orbits, and Origin of Comets. Proceedings of 45$^{th}$ IAU
  Symposium}, pp.\ 55-60. Springer-Verlag, New York/Berlin.

\bibitem[Rees(1988)]{rees:88} Rees, M. 1988, Nature 333, 523.

\bibitem[Rosswog \& Trautman(1996)]{rosswog:96} Rosswog, S., \&
  Trautman, D. 1996, Plant.\ Space Sci.\ 44, 313.

\bibitem[Salo \& Yoder(1988)]{salo:88} Salo, H., \& Yoder, C.\
  F. 1988, A\& A 205, 309.

\bibitem[Schnittman(2004)]{schnittman:04} Schnittman, J.\ D. 2004, Phys.\
Rev.\ D, 124020.

\bibitem[Seto \& Muto(2010)]{seto:10} Seto, N., \& Muto, T. 2010,
  Phys.\ Rev.\ D, in press. [arXiv:1005.3114].

\bibitem[Shakura \& Sunyaev(1973)]{shakura:73} Shakura, N.\ I., \& Sunyaev,
  R.\ A. 1973, A\&A, 24, 337

\bibitem[Shoemaker et al.(1989)]{shoemaker:89} Shoemaker, E.\ M.,
  Shoemaker, C.\ S., \& Wolfe, R.\ F. 1989, in {\it Asteroids II},
  pp.\ 487-523. Univ.\ of Arizona Press, Tucson.

\bibitem[Stone \& Loeb(2010)]{stone:10} Stone, N., \& Loeb, A. 2010,
  [arXiv:1004.4833]. 

\bibitem[Ulmer(1999)]{ulmer:99} Ulmer, A. 1999, ApJ 514, 180.

\bibitem[van Meter et al.(2010)]{vanmeter:10} van Meter, J.\ R., Wise,
  J.\ H., Miller, M.\ C., Reynolds, C.\ S., Centrella, J., Baker, J.\
  G., Boggs, W.\ D., Kelly, B.\ J., \& McWilliams, S.\ T. 2010, ApJ 711,
  L89. 

\bibitem[Wang \& Merritt(2004)]{wang:04} Wang, J., \& Merritt,
  D. 2004, ApJ 600, 149.

\bibitem[Yoder(1979)]{yoder:79} Yoder, C.\ F. 1979, Icarus 40, 341.

\end{thebibliography}
\end{document}